\documentclass[pra,onecolumn,aps,superscriptaddress,showpacs,showkeys,floatfix]{revtex4}
\usepackage{amssymb}
\usepackage{graphicx}
\usepackage{dcolumn}
\usepackage{bm}
\usepackage{amsmath}
\usepackage{amsbsy}
\usepackage{amssymb}
\usepackage{graphicx}
\usepackage{color}
\usepackage{epstopdf}
\usepackage{mathrsfs}
\usepackage{multirow}
\usepackage[colorlinks,linkcolor=magenta,citecolor=blue,urlcolor=blue]{hyperref}
\usepackage{etoolbox}
\makeatletter
\patchcmd{\@makecaption}
  {\scshape}
  {}
  {}
  {}
\makeatother

\makeatletter
\newcommand{\Rmnum}[1]{\expandafter\@slowromancap\romannumeral #1@}
\makeatother

\begin{document}

\title{Generalized gauge transformation with $PT$-symmetric non-unitary
operator and classical correspondence of non-Hermitian Hamiltonian for a
periodically driven system}
\author{Yan Gu}
\author{Xiao-Lei Hao}
\author{J.-Q. Liang}
\email{jqliang@sxu.edu.cn}
\affiliation{Institute of Theoretical Physics and Department of Physics, State Key
Laboratory of Quantum Optics and Quantum Optics Devices, Shanxi University,
Taiyuan, Shanxi 030006, China.}

\begin{abstract}
We in this paper demonstrate that the $PT$-symmetric non-Hermitian
Hamiltonian for a periodically driven system can be generated from a kernel
Hamiltonian by a generalized gauge transformation. The kernel Hamiltonian is
Hermitian and static, while the time-dependent transformation operator has
to be $PT$ symmetric and non-unitary in general. Biorthogonal sets of
eigenstates appear necessarily as a consequence of non-Hermitian
Hamiltonian. We obtain analytically the wave functions and associated
non-adiabatic Berry phase $\gamma _{n}$ for the $n$th eigenstate. The
classical version of the non-Hermitian Hamiltonian becomes a complex
function of canonical variables and time. The corresponding kernel
Hamiltonian is derived with $PT$ symmetric canonical-variable transfer in
the classical gauge transformation. Moreover, with the change of
position-momentum to angle-action variables it is revealed that the
non-adiabatic Hannay's angle $\Delta \theta _{H}$ and Berry phase satisfy
precisely the quantum-classical correspondence, $\gamma _{n}=$ $%
(n+1/2)\Delta \theta _{H}$.
\end{abstract}

\date{[]}
\maketitle

\volumeyear{year} \volumenumber{number} \issuenumber{number} \eid{identifier}

\received[Received text]{date}

\revised[Revised text]{date}

\accepted[Accepted text]{date}

\published[Published text]{date}

\startpage{1} \endpage{2} \preprint{ }

%\accepted{(Day Month Year)}
%\comby{(xxxxxxxxxx)}

\section{Introduction}

In standard quantum theory, observable quantities are associated with
Hermitian operators, and time evolution is generated by a Hermitian
Hamiltonian{\textsuperscript{\cite{Croke2015}}}. The Hermiticity (or more
precisely self-adjointness {\textsuperscript{\cite{Lax2002}}}) of the
Hamiltonian ensures both the reality of the energy spectrum and more
importantly the unitarity of time evolution. Surprisingly, Bender and
Boettcher showed in 1998 that the non-Hermitian Hamiltonians can still
possess real and positive eigenvalues{\textsuperscript{\cite{Bender1998}}}.
They claimed that the reality of the spectrum is due to parity-time ($PT$)
symmetry of Hamiltonian. The appearance of the $PT$-symmetry was initially
considered as a mathematically interesting finding{\textsuperscript{%
\cite{Bender1998,Bender2002,Bender2007}}}, with which many new Hamiltonians
can be constructed instead of the Hermiticity{\textsuperscript{%
\cite{Bender2005}}}. The predicted properties of $PT$-symmetric Hamiltonians{%
\textsuperscript{\cite{Bender2007,Most2010}}} have been observed at the
classical level in a wide variety of laboratory experiments involving
superconductivity{\textsuperscript{\cite{Rubi2007,Chtc2012}}}, optics{%
\textsuperscript{\cite{Guo2009,Ruter2010,Lin2011,Feng2011}}}, microwave
cavities{\textsuperscript{\cite{Bittner2012}}}, atomic diffusion{%
\textsuperscript{\cite{Zhao2010}}}, nuclear magnetic resonance{%
\textsuperscript{\cite{Zheng2013}}}. F. Klauck $et$ $al$. demonstrate
two-particle quantum interference in a $PT$-symmetric system{%
\textsuperscript{\cite{Klauck2019}}}.

The connection between $PT$ symmetry and positive spectra is simply
illustrated by adding a $PT$-symmetric non-Hermitian term to the Hamiltonian
of a harmonic oscillator {\textsuperscript{\cite{Bender1998}}}. If the
harmonic oscillator is extended into the complex domain more intricate
behavior is discovered{\textsuperscript{\cite{Bender2019}}}. The $PT$%
-symmetric oscillator has been investigated extensively in recent years. In
2005, Cem Yuce studied the exactly solvable $PT$ symmetric harmonic
oscillator problem{\textsuperscript{\cite{Yuce2005}}}. In 2013, Carl M.
Bender $et$ $al$. observed\ $PT$ phase transition in a two-oscillator model{%
\textsuperscript{\cite{Benderajp}}}, and studied twofold transitions as well{%
\textsuperscript{\cite{Bender2013}}}. Two years later, Alireza Beygi $et$ $%
al $. examined coupled oscillator chain with partial $PT$ symmetry{%
\textsuperscript{\cite{Beygi2015}}}. Subsequently Andreas Fring and Thomas
Frith presented exact analytical solutions for a two-dimensional
time-dependent non-Hermitian system by coupling two harmonic oscillators
with infinite dimensional Hilbert space in the broken $PT$-regime{%
\textsuperscript{\cite{Fring2018}}}. And then they provided a time-dependent
Dyson map and metric for the two-dimensional, non-Hermitian oscillator in
2020{\textsuperscript{\cite{Fring2020}}}. The dynamics of the average
displacement of the mechanical oscillator were investigated in different
regimes for the $PT$-symmetric, optomechanical system in 2021{%
\textsuperscript{\cite{Xu2021}}}.

In the present paper we study the gauge equivalence of a $PT$ symmetric
Hamiltonian with a Hermitian Hamiltonian for a periodically driven system.
Since the Schr\"{o}dinger equation is invariant under the $PT$
transformation for $PT$ symmetric non-Hermitian Hamiltonian, we can generate
various Hamiltonians by means of the $PT$ symmetric but non-unitary
transformation operators, different from the ordinary quantum mechanics, in
which the transformation operators have to be unitary. Moreover classical
correspondence of $PT$-symmetric non-Hermitian is formulated explicitly in
the viewpoint of gauge equivalence. The quantum--classical correspondence
was demonstrated for the first time by Schr\"{o}dinger that the center of
wave packet in coherent state evolves in time following exactly the
corresponding motion of classical particle{\textsuperscript{\cite{Schr1926}}}%
. In fact the quantum-classical correspondence has resulted in many
impressive aspects showing the unite nature of quantum and classical worlds.
For two dimensional harmonic-oscillator, the quantum wave packet of coherent
states corresponds exactly to the transformation geometry of classical
dynamics{\textsuperscript{\cite{Chen2011}}}. Y. F. Chen $et$ $al$.
manifested the connection between topological structures of quantum
stationary coherent-states and bundles of classical Lissajous orbits in 2018{%
\textsuperscript{\cite{Chen2018}}}. V. Gattus and S. Karamitsos show that
the uncertainties of certain coupled classical-systems and their quantum
counterparts converge in the classical limit{\textsuperscript{%
\cite{Gattus2020}}}. Quantum-classical correspondence is also demonstrated
by comparing the statistical moments of eigenfunctions in closed systems
with chaotic dynamics{\textsuperscript{\cite{Saldana2021}}}.

When a classical or quantum system undergoes a cyclic evolution governed by
slow change in its parameter space, it acquires a topological phase factor
known as the geometric or Berry phase{\textsuperscript{\cite{Berry1984}}},
which reveals the gauge structure in quantum mechanics{\textsuperscript{%
\cite{Berry1984}}}. Hannay's angle is the classical counterpart of this
additional quantum phase{\textsuperscript{\cite{Hannay1985,Berry1985}}}. The
quantum geometric-phase and the classical Hannay's angle come, as a matter
of fact, from the same nature in the semiclassical level according to Berry{%
\textsuperscript{\cite{Berry1985}}}. Recently, L. Latmiral and F. Armata
addressed the quantum-classical comparison of phase measurements in
optomechanics for composite systems{\textsuperscript{\cite{Latmiral2020}}}.
A. R\"{u}ckriegel and R. A. Duine study the geometric phases that occur in
semi-classical magnetic dynamics{\textsuperscript{\cite{Ruckriegel2020}} }%
within the framework developed by Hannay for classical integrable systems.

We propose a gauge transformation method to solve the time-dependent
non-Hermitian Hamiltonian with $PT$ symmetric but non-unitary operator.
Quantum wave functions are obtained analytically along with the
non-adiabatic Berry phase for the periodically driven system. Then classical
counterpart of the non-Hermitian Hamiltonian is solved in terms of the gauge
transformation in classical mechanics. The explicit relation of Hannay's
angle and Berry phase is presented for the $PT$-symmetric non-Hermitian
Hamiltonian.

The paper is organized as follows. We explore in Sec. II the $PT$-symmetric
non-Hermitian Hamiltonian consisting of periodically driven $SU(1,1)$
generators. The kernel Hamiltonian, which is Hermitian and static is derived
by means of generalized gauge transformation with $PT$-symmetric non-unitary
operator. The biorthogonal sets of eigenfunctions are obtained along with
associated Berry phases. In Sec. III, we present the corresponding variation
of classical Hamiltonian in the gauge transformation of classical mechanics.
And the Hannay's angle is found to have the precise correspondence with the
Berry phase{\textsuperscript{\cite{Liu1998,Xin2015}}}. The conclusion is
given in the last section.

\section{ $PT$-symmetric non-Hermitian Hamiltonian and non-unitary
transformation operator}
The $PT$-symmetric non-Hermitian Hamiltonian that we consider is written as%
\begin{equation}
\hat{H}\left( t\right) =\Omega \hat{S}_{z}+G\left( \hat{S}_{+}e^{i\phi
\left( t\right) }-\hat{S}_{-}e^{-i\phi \left( t\right) }\right)  \label{h}
\end{equation}%
in which $\phi \left( t\right) =$ $\omega t$ with $\omega $ being the
driving frequency, $\Omega \ $and$\ G$ are the frequency parameters. The $%
SU(1,1)$ generators $\hat{S}_{z}$ and $\hat{S}_{+}=\left( \hat{S}_{-}\right)
^{\dag }$ satisfy the communication relation{\textsuperscript{%
\cite{Abdalla2017}}}%
\begin{equation}
\left[ \hat{S}_{z},\hat{S}_{\pm }\right] =\pm \hat{S}_{\pm },\quad \left[
\hat{S}_{+},\hat{S}_{-}\right] =-2\hat{S}_{z},  \label{c}
\end{equation}%
where natural unit $\hbar =1$ is used throughout. The Hamiltonian Eq.(\ref{h}%
) is obviously non-Hermitian {\textsuperscript{\cite{Bender2005}}}%
\begin{equation*}
\hat{H}\left( t\right) \neq \hat{H}^{\dag }\left( t\right) =\Omega \hat{S}%
_{z}-G\left( \hat{S}_{+}e^{i\phi \left( t\right) }-\hat{S}_{-}e^{-i\phi
\left( t\right) }\right) .
\end{equation*}%
The $PT$ transformation operator in quantum mechanics can be defined as
\begin{equation}
\hat{\Theta}=\hat{U}K,\quad \hat{\Theta}^{-1}=K\hat{U}^{\dag }  \label{pt}
\end{equation}%
where $\hat{U}$ is the usual unitary operator for the space-time reflection
and $K$ denotes the operation of complex conjugate. The position and
momentum operators become $\hat{\Theta}\hat{x}\hat{\Theta}^{-1}=-\hat{x}$, $%
\hat{\Theta}\hat{p}\hat{\Theta}^{-1}=\hat{p}$ under $PT$ transformation. And
the boson creation operator $\hat{a}^{\dag }=\frac{1}{\sqrt{2}}\left( \hat{x}%
-i\hat{p}\right) $ and annihilation operator $\hat{a}=\frac{1}{\sqrt{2}}%
\left( \hat{x}+i\hat{p}\right) $ in harmonic oscillator become%
\begin{equation*}
\hat{\Theta}\hat{a}^{\dag }\hat{\Theta}^{-1}=-\hat{a}^{\dag },\hat{\Theta}%
\hat{a}\hat{\Theta}^{-1}=-\hat{a}.
\end{equation*}%
The $SU(1,1)$ Lie algebra has a realization in terms of boson creation and
annihilation operators $\hat{a}^{\dag }$\ and $\hat{a}$ such that{%
\textsuperscript{\cite{Liang2020}}}%
\begin{equation}
\hat{S}_{z}=\frac{1}{2}\left( \hat{a}^{\dag }\hat{a}+\frac{1}{2}\right) ,%
\hat{S}_{+}=\frac{1}{2}\left( \hat{a}^{\dag }\right) ^{2},\hat{S}_{-}=\frac{1%
}{2}\left( \hat{a}\right) ^{2}.  \label{su}
\end{equation}%
Thus, under the $PT$ transformation the $SU(1,1)$ generators transform as%
\begin{equation}
\hat{\Theta}\hat{S}_{z}\hat{\Theta}^{-1}=\hat{S}_{z},\ \hat{\Theta}\hat{S}%
_{+}\hat{\Theta}^{-1}=\hat{S}_{+},\ \hat{\Theta}\hat{S}_{-}\hat{\Theta}^{-1}=%
\hat{S}_{-}.
\end{equation}%
The commutation relation Eq.(\ref{c}) is $PT$-transformation invariant as
well as the Hamiltonian{\textsuperscript{\cite{Bender2005}}}
\begin{equation*}
\hat{\Theta}\hat{H}\left( t\right) \hat{\Theta}^{-1}=\hat{H}\left( t\right) .
\end{equation*}%
The time-dependent Schr\"{o}dinger equation is covariant under $PT$
transformation
\begin{equation*}
i\frac{\partial }{\partial t}\left\vert \psi ^{\prime }\left( t\right)
\right\rangle =\hat{H}\left( t\right) \left\vert \psi ^{\prime }\left(
t\right) \right\rangle ,
\end{equation*}%
where $\left\vert \psi ^{\prime }\left( t\right) \right\rangle =\hat{\Theta}%
\left\vert \psi \left( t\right) \right\rangle $.

\subsection{Generalized gauge transformation with non-unitary operator}

Since the Schr\"{o}dinger equation is invariant under $PT$ transformation,
it is always possible to construct a proper transformation such that the
Hamiltonian in the new gauge can be easily diagonalized. We propose a gauge
transformation method to solve the time-dependent non-Hermitian Hamiltonian
system with a transformation operator considered as%
\begin{eqnarray}
\hat{R}\left( t\right) &=&e^{-\frac{\eta }{2}\left( \hat{S}_{+}e^{i\phi
\left( t\right) }+\hat{S}_{-}e^{-i\phi \left( t\right) }\right) },  \label{r}
\\
\hat{R}^{-1}\left( t\right) &=&e^{\frac{\eta }{2}\left( \hat{S}_{+}e^{i\phi
\left( t\right) }+\hat{S}_{-}e^{-i\phi \left( t\right) }\right) },  \notag
\end{eqnarray}%
in which the real parameter $\eta $ is to be determined. The operator Eq.(%
\ref{r}) is $PT$-symmetric
\begin{equation*}
\hat{\Theta}\hat{R}\left( t\right) \hat{\Theta}^{-1}=\hat{R}\left( t\right) ,
\end{equation*}%
in order to maintain the $PT$ symmetry of the transformed Hamiltonian.
However the operator $\hat{R}\left( t\right) $ is non-unitary with $\hat{R}%
^{\dag }\neq \hat{R}^{-1}$ different from the ordinary quantum mechanics, in
which the transformation operator has to be unitary. Occasionally it is
Hermitian ($\hat{R}^{\dag }(t)=\hat{R}(t)$) in this particular model but is
unnecessary in general. The time-dependent Schr\"{o}dinger equation becomes
\begin{equation}
i\frac{\partial }{\partial t}\left\vert \psi ^{\prime }\left( t\right)
\right\rangle =\hat{H}^{\prime }\left( t\right) \left\vert \psi ^{\prime
}\left( t\right) \right\rangle .  \label{schr}
\end{equation}%
where $\hat{R}\left( t\right) \left\vert \psi \left( t\right) \right\rangle
=\left\vert \psi ^{\prime }\left( t\right) \right\rangle $. The Hamiltonian
in the new gauge is%
\begin{equation}
\hat{H}^{\prime }\left( t\right) =\hat{R}\left( t\right) \hat{H}\left(
t\right) \hat{R}^{-1}\left( t\right) -i\hat{R}\left( t\right) \frac{\partial
}{\partial t}\hat{R}^{-1}\left( t\right) .  \label{newh}
\end{equation}

With the help of $SU(1,1)$ commutation relation Eq.(\ref{c}) and the
Baker-Hausdorff-Campbell formula we have the following transformations
\begin{eqnarray*}
\hat{R}\left( t\right) \hat{S}_{+}\hat{R}^{-1}\left( t\right) &=&\hat{S}%
_{+}\cos ^{2}\left( \frac{\eta }{2}\right) -\hat{S}_{z}e^{-i\phi \left(
t\right) }\sin \left( \eta \right) \\
&&+\hat{S}_{-}e^{-2i\phi \left( t\right) }\sin ^{2}\left( \frac{\eta }{2}%
\right) \\
\hat{R}\left( t\right) \hat{S}_{-}\hat{R}^{-1}\left( t\right) &=&\hat{S}%
_{-}\cos ^{2}\left( \frac{\eta }{2}\right) +\hat{S}_{z}e^{i\phi \left(
t\right) }\sin \left( \eta \right) \\
&&+\hat{S}_{+}e^{2i\phi \left( t\right) }\sin ^{2}\left( \frac{\eta }{2}%
\right) \\
\hat{R}\left( t\right) \hat{S}_{z}\hat{R}^{-1}\left( t\right) &=&\hat{S}%
_{z}\cos \left( \eta \right) +\frac{1}{2}\sin \left( \eta \right) \left(
\hat{S}_{+}e^{i\phi \left( t\right) }-\hat{S}_{-}e^{-i\phi \left( t\right)
}\right)
\end{eqnarray*}%
and%
\begin{eqnarray}
i\hat{R}\left( t\right) \frac{\partial }{\partial t}\hat{R}^{-1}\left(
t\right) &=&2\frac{d\phi }{dt}\hat{S}_{z}\sin ^{2}\left( \frac{\eta }{2}%
\right)  \notag \\
&&-\frac{d\phi }{2dt}\sin \left( \eta \right) \left( \hat{S}_{+}e^{i\phi
\left( t\right) }-\hat{S}_{-}e^{-i\phi \left( t\right) }\right)  \label{NR}
\end{eqnarray}

Substituting transformation relations Eq.(\ref{NR}) into the Hamiltonian Eq.(%
\ref{newh}) in the new gauge yields%
\begin{eqnarray*}
\hat{H}^{\prime } &=&\left[ \Omega \cos \left( \eta \right) -2G\sin \left(
\eta \right) -2\omega \sin ^{2}\left( \frac{\eta }{2}\right) \right] \hat{S}%
_{z} \\
&&+\left[ G\cos \left( \eta \right) +\frac{\Omega +\omega }{2}\sin \left(
\eta \right) \right] \left( \hat{S}_{+}e^{i\phi \left( t\right) }-\hat{S}%
_{-}e^{-i\phi \left( t\right) }\right) ,
\end{eqnarray*}%
which is $PT$ symmetric. The parameter $\eta $ can be determined from the
following auxiliary equation
\begin{equation*}
G\cos \left( \eta \right) +\frac{\omega +\Omega }{2}\sin \left( \eta \right)
=0,
\end{equation*}%
which leads to
\begin{equation*}
\sin \left( \eta \right) =\pm \frac{2G}{\Delta },
\end{equation*}%
with
\begin{equation*}
\Delta =\sqrt{\left( \omega +\Omega \right) ^{2}+4G^{2}}.
\end{equation*}%
The Hamiltonian in new gauge Eq.(\ref{newh}) then becomes time independent
and Hermitian
\begin{equation}
\hat{H}^{\prime }=2\Gamma \hat{S}_{z},  \label{new}
\end{equation}%
with an effective frequency $\Gamma $%
\begin{equation*}
\Gamma =\pm \frac{\left( \omega +\Omega \right) ^{2}-4G^{2}}{2\Delta }-\frac{%
\omega }{2}.
\end{equation*}%
Let $\left\vert n\right\rangle $ be the eigenstate of $2\hat{S}_{z}=\left(
\hat{a}^{\dag }\hat{a}+1/2\right) $ such that $\hat{a}^{\dag }\hat{a}%
\left\vert n\right\rangle =n\left\vert n\right\rangle $, the energy
eigenvalues of $\hat{H}^{\prime }$ are that of a simple Harmonic oscillator
\begin{equation*}
E_{n}=\left( n+\frac{1}{2}\right) \Gamma .
\end{equation*}%
The\ $PT$-symmetric non-Hermitian Hamiltonian has pure real-eigenvalues
known as the unbroken $PT$ symmetry{\textsuperscript{\cite{Benderajp}}}. We
have shown that the $PT$-symmetric non-Hermitian Hamiltonian can be
transformed into a Hermitian one in terms of the generalized gauge
transformation.

Recently, it was demonstrated that a PT-symmetric non-Hermitian Hamiltonian
can be transformed into the corresponding Hermitian one with the vielbein
formalism. The generalized gauge transformation could be a special case of
the vielbein formalism\textsuperscript{\cite{Ju1}}.

\subsection{Biorthogonal sets of eigenstates and non-adiabatic Berry phase}

The special solution of the Schr\"{o}dinger equation in the original gauge
for the eigenstate $\left\vert n\right\rangle $ is
\begin{equation}
\left\vert \psi _{n}\left( t\right) \right\rangle _{r}=e^{-iE_{n}t}\hat{R}%
^{-1}\left( t\right) \left\vert n\right\rangle ,  \label{k}
\end{equation}%
in which the subscript "$r$" denotes "ket" state since these states resulted
from non-unitary transformation are not orthonormal. The corresponding "bra"
state is defined by%
\begin{equation*}
\left\vert \psi _{n}\left( t\right) \right\rangle _{l}=e^{-iE_{n}t}\hat{R}%
\left( t\right) \left\vert n\right\rangle .
\end{equation*}%
The orthonormal condition is%
\begin{equation}
_{l}\langle \psi _{n}\left( t\right) |\psi _{m}\left( t\right) \rangle
_{r}=\delta _{nm}.  \label{o}
\end{equation}%
According to Ref.\textsuperscript{\cite{Ju2}}, we can define a metric
operator $\hat{\chi}$ relating the "ket" and "bra" states by%
\begin{equation}
\left\vert \psi _{n}(t)\right\rangle _{l}=\hat{\chi}\left\vert \psi
_{n}(t)\right\rangle _{r},  \label{mt}
\end{equation}%
in which the metric operator of Hilbert space for the present model is%
\begin{equation}
\hat{\chi}=\hat{R}^{2}\left( t\right) .  \label{m}
\end{equation}%
With the metric operator orthogonality condition Eq.(\ref{o}) becomes
\begin{equation}
\langle \langle \psi _{n}(t)|\psi _{m}(t)\rangle \rangle \equiv \left\langle
\psi _{n}(t)\right\vert \hat{\chi}\left\vert \psi _{m}(t)\right\rangle
=\delta _{nm}  \label{G}
\end{equation}%
without using the two sets of basis. The equation of motion for the metric
operator $\hat{\chi}$ is given explicitly in Ref.\textsuperscript{\cite{Ju2}}
for a full description of the non-Hermitian system dynamics.

The time dependent state Eq.(\ref{k}) can be rewritten as%
\begin{equation*}
\left\vert \psi _{n}\left( t\right) \right\rangle
_{r}=e^{-i\int_{0}^{t}\left\langle n\right\vert \hat{R}\hat{H}\hat{R}^{-1}-i%
\hat{R}\frac{\partial }{\partial t}\hat{R}^{-1}\left\vert n\right\rangle dt}%
\hat{R}^{-1}\left( t\right) \left\vert n\right\rangle ,
\end{equation*}%
from which the non-adiabatic Berry phase in one period ($T=2\pi /\omega $)
of the driving field is evaluated by
\begin{equation*}
\gamma _{n}\left( T\right) =i\int_{0}^{T}\left\langle n\right\vert \hat{R}%
\frac{\partial }{\partial t}\hat{R}^{-1}\left\vert n\right\rangle dt.
\end{equation*}%
With the transformation operator $\hat{R}$ the explicit form of Berry phase
is found as
\begin{equation}
\gamma _{n}\left( T\right) =\pi \left( n+\frac{1}{2}\right) \left( 1\mp
\frac{\omega +\Omega }{\Delta }\right) .  \label{berry}
\end{equation}

As a matter of fact the Hamiltonian Eq.(\ref{new}) becomes in the
representation of coordinate and momentum operators
\begin{equation}
\hat{H}^{\prime }=\frac{\Gamma }{2}\left( \hat{X}^{2}+\hat{P}^{2}\right) ,
\label{q}
\end{equation}%
in which $\hat{X}=\frac{1}{\sqrt{2}}(\hat{a}+\hat{a}^{\dag })$ and $\hat{P}=%
\frac{1}{\sqrt{2}i}(\hat{a}-\hat{a}^{\dag })$ as usual. The eigenstate
wave-functions are given by%
\begin{equation*}
\psi _{n}^{\prime }\left( X,\Gamma \right) =\sqrt{\frac{\Gamma ^{\frac{1}{2}}%
}{2^{n}\pi ^{\frac{1}{2}}n!}}e^{-\frac{\Gamma X}{2}}H_{n}\left( \Gamma ^{%
\frac{1}{2}}X\right) ,
\end{equation*}%
where $H_{n}(\Gamma ^{\frac{1}{2}}X)$ is the Hermit polynomial expressed as%
\begin{equation*}
H_{n}\left( \Gamma ^{\frac{1}{2}}X\right) =\left( -1\right) ^{n}e^{\Gamma
X^{2}}\frac{d^{n}}{d\left( \Gamma ^{\frac{1}{2}}X\right) ^{n}}e^{-\Gamma
X^{2}}.
\end{equation*}%
The special solution of the time-dependent Schr\"{o}dinger equation (\ref%
{schr}) for $n$th eigenvalue is obviously
\begin{equation*}
\psi _{n}^{\prime }\left( X,\Gamma ,t\right) =\psi _{n}^{\prime }\left(
X,\Gamma \right) e^{-iE_{n}t}.
\end{equation*}

The wave function of original Schr\"{o}dinger equation for the
time-dependent Hamiltonian $\hat{H}\left( t\right) $ Eq.(\ref{h}) can be
obtained from $\psi _{n}^{\prime }\left( X,\Gamma ,t\right) $ with the
inverse transformation operator such that%
\begin{equation*}
\psi _{n}\left( X,\Gamma ,t\right) =\hat{R}^{-1}\left( X,P,t\right) \psi
_{n}^{\prime }\left( X,\Gamma ,t\right) ,
\end{equation*}%
where the transformation operator in the coordinate representation is given
by%
\begin{equation*}
\hat{R}^{-1}\left( t\right) =e^{\frac{\eta }{4}\left[ \left( X^{2}+\frac{%
\partial ^{2}}{\partial X^{2}}\right) \cos \phi \left( t\right) -i\left( 2X%
\frac{\partial }{\partial X}+1\right) \sin \phi \left( t\right) \right] }.
\end{equation*}

\section{Classical counterpart of non-Hermitian Hamiltonian and canonical
transformation}

The Hannay's angle is a classical analogue of the quantum Berry phase{%
\textsuperscript{\cite{Berry1985,Liu2011}}}. Therefore, it is of interest to
see the classical correspondence of the non-Hermitian Hamiltonian and the
related Berry phase. In terms of the boson operator representation of the $%
SU(1,1)$ generators Eq.(\ref{su}), the classical version of the $PT$%
-symmetric non-Hermitian Hamiltonian Eq.(\ref{h})\ becomes%
\begin{eqnarray}
H\left( x,p,t\right) &=&\left[ \frac{\Omega }{4}+i\frac{G}{2}\sin \phi
\left( t\right) \right] x^{2}+\left[ \frac{\Omega }{4}-i\frac{G}{2}\sin \phi
\left( t\right) \right] p^{2}  \notag \\
&&-iG\cos \phi \left( t\right) xp.  \label{class}
\end{eqnarray}%
Surprisingly, but naturally, the classical Hamiltonian is a complex function
of canonical variable $x$, $p$ and time $t$. It is obviously $PT$ symmetric,
noticing that the $PT$ transformation operator Eq.(\ref{pt}) is also valid
in the classical mechanics including the complex conjugation operation.

\subsection{Time-dependent canonical transformation}

The classical kernel Hamiltonian can be found with the help of a
time-dependent canonical transformation{\textsuperscript{\cite{Xin2015}}}
corresponding to the quantum transformation Eq.(\ref{r})
\begin{eqnarray}
x &=&\left[ \cos \frac{\eta }{2}-i\alpha \sin \frac{\eta }{2}\right]
X+i\beta P\sin \frac{\eta }{2},  \notag \\
p &=&i\beta X\sin \frac{\eta }{2}+\left[ \cos \frac{\eta }{2}+i\alpha \sin
\frac{\eta }{2}\right] P,  \label{tran}
\end{eqnarray}%
in which $\alpha =\sin \phi \left( t\right) $, $\beta =\cos \phi \left(
t\right) $ and the parameter $\eta $ is to be determined. The above
transformation Eq.(\ref{tran}) is obviously $PT$ symmetric.

It is well known that two Lagrangians, which differ by a total
time-derivative of space-time function, give rise to the same equation of
motion. This is called the generalized gauge transformation in classical
mechanics{\textsuperscript{\cite{Liang2020}}}. We require that the
phase-space Lagrangians{\textsuperscript{\cite{Liang2020}}}%
\begin{equation*}
L^{\prime }\left( X,P\right) =P\dot{X}-H^{\prime }\left( X,P,t\right)
\end{equation*}%
in new variables and%
\begin{equation*}
L\left( x,p\right) =p\dot{x}-H\left( x,p,t\right)
\end{equation*}%
in the original variables differ by a total time-derivative of the
generating function%
\begin{equation*}
L^{\prime }\left( X,P\right) =L\left( x,p\right) +\frac{dF\left(
X,P,t\right) }{dt}.
\end{equation*}%
So that $L^{\prime }\left( X,P\right) $ and $L\left( x,p\right) $ are gauge
equivalent. The generating function is constructed as%
\begin{eqnarray*}
F\left( X,P,t\right) &=&\left( -i\frac{\beta }{2}\sin \eta -\alpha \beta
\sin ^{2}\frac{\eta }{2}\right) \frac{X^{2}}{2} \\
&&+\beta ^{2}XP\sin ^{2}\frac{\eta }{2} \\
&&+\left( -i\frac{\beta }{2}\sin \eta +\alpha \beta \sin ^{2}\frac{\eta }{2}%
\right) \frac{P^{2}}{2}.
\end{eqnarray*}%
We choose the parameter value $\eta $ such that%
\begin{equation*}
\sin \eta =-\frac{2G}{\Delta }
\end{equation*}%
the Hamiltonian in the new gauge becomes exactly the classical counterpart
of the quantum version Eq.(\ref{q})
\begin{equation}
H^{\prime }\left( X,P\right) =\frac{\Gamma }{2}\left( X^{2}+P^{2}\right) .
\label{newH}
\end{equation}

\subsection{Action-angle variable and Hannay's angle}

If an integrable classical-Hamiltonian $H$ describing a bound motion depends
on slowly varying parameters the action variable $I$ of the motion is
conserved according to Hannay{\textsuperscript{\cite{Hannay1985}}}. The
frequency of quasi-periodic motion $\Gamma =\partial H^{\prime }(I,\Theta
)/\partial I$ does not vanish, and then the action variable $I$ is an
adiabatic invariant. According to the canonical equations%
\begin{eqnarray*}
\dot{\Theta} &=&\frac{\partial H^{\prime }(I,\Theta )}{\partial I}=\Gamma ,
\\
\dot{I} &=&-\frac{\partial H^{\prime }(I,\Theta )}{\partial \Theta }=0,
\end{eqnarray*}%
the Hamiltonian $H^{\prime }\left( X,P\right) $ becomes in the action-angle
variables%
\begin{equation*}
H^{\prime }(I,\Theta )=I\Gamma .
\end{equation*}%
The Hannay's angle in classical mechanics is introduced with the canonical
transformation from the position-momentum variables $(X,P)$ to the
action-angle variables $(I,\Theta )$ such that%
\begin{equation}
X(I,\Theta )=\sqrt{2I}\sin \Theta ,\text{ }P(I,\Theta )=\sqrt{2I}\cos \Theta
.  \label{act}
\end{equation}

Substituting the $X(I,\Theta )$ and $P(I,\Theta )$ defined in Eq.(\ref{act})
into Eq.(\ref{tran}), we have the canonical transformation from the original
variables $(x,p)$ to the action-angle variables $(I,\Theta )$,%
\begin{eqnarray*}
x(I,\Theta ) &=&\sqrt{2I}\left[ \left( \cos \frac{\eta }{2}-i\alpha \sin
\frac{\eta }{2}\right) \sin \Theta +i\beta \sin \frac{\eta }{2}\cos \Theta %
\right] , \\
p(I,\Theta ) &=&\sqrt{2I}\left[ i\beta \sin \frac{\eta }{2}\sin \Theta
+\left( \cos \frac{\eta }{2}+i\alpha \sin \frac{\eta }{2}\right) \cos \Theta %
\right] .
\end{eqnarray*}%
The Hannay's angle of the time-dependent system with the Hamiltonian $%
H\left( x,p,t\right) $ in the original gauge is evaluated in one-period $%
T=2\pi /\omega $ as{\textsuperscript{\cite{Berry1985}}}
\begin{eqnarray}
\Delta \theta _{H} &=&-\frac{\partial }{\partial I_{c}}\left%
\langle p\left( I,\Theta \right) d_{\phi }x\left( I,\Theta \right)
\right\rangle _{\Theta }  \notag \\
&=&-\frac{\partial }{\partial I}\frac{1}{2\pi }\int_{0}^{2\pi }d\Theta
\int_{0}^{2\pi }p\left( I,\Theta \right) d_{\phi }x\left( I,\Theta \right)
\notag \\
&=&\pi \left( \pm \frac{\omega +\Omega }{\Delta }-1\right)  \label{hannay}
\end{eqnarray}%
where $\left\langle ...\right\rangle _{\Theta }$ denotes the average over
the angle variable $\Theta $.

We reveal a precise relation{\textsuperscript{\cite{Liu1998,Xin2015}}}%
\begin{equation*}
\gamma _{n}\left( T\right) =-\left( n+1/2\right) \Delta \theta _{H}
\end{equation*}%
between the non-adiabatic Berry phase Eq.(\ref{berry}) and Hannay angle Eq.(%
\ref{hannay}) for the $n$th eigenstates.

\section{Conclusion}

The time-dependent Schr\"{o}dinger equation is invariant under the $PT$
transformation for the $PT$-symmetric non-Hermitian Hamiltonian. Based on
the invariance we propose a generalized gauge transformation with $PT$%
-symmetric, however non-unitary operator different from the ordinary quantum
mechanics. It is demonstrated that the non-Hermitian Hamiltonian can be
generated from a kernel Hamiltonian, which is Hermitian and static, by the
gauge transformation. Thus the time-dependent non-Hermitian Hamiltonian is
solved analytically. We obtain explicit quantum wave-functions $\psi
_{n}\left( X,\Gamma ,t\right) $ and associated nonadiabatic Berry phase $%
\gamma _{n}$. The classical counterpart of the Non-Hermitian Hamiltonian can
be transformed by the classical gauge transformation to an exactly same form
as the quantum Kernel-Hamiltonian. The quantum-classical correspondence is
then established explicitly for the $PT$-symmetric non-Hermitian Hamiltonian
system. Moreover the classical Hannay's angle $\Delta \theta _{H}$ is
derived by the canonical-variable transfer to action-angle variables. We
reveal a precise relation of quantum Berry phase $\gamma _{n}$ and classical
Hannay angle $\Delta \theta _{H}$.

\section*{Acknowledgements}

This work was supported by the National Natural Science Foundation of China
(Grant Nos. 11275118 and 11874246).

\section*{Conflict of Interest}

The authors declare no conflict of interest.

\end{document}